# MULTI-MODALITY IMAGE SUPER-RESOLUTION USING GENERATIVE ADVERSARIAL NETWORKS

Aref Abedjooy
*Faculty of Science, Ontario Tech University*
*Oshawa, Ontario, Canada*
aref.abedjooydivshali@ontariotechu.net

Mehran Ebrahimi
*Faculty of Science, Ontario Tech University*
*Oshawa, Ontario, Canada*
mehran.ebrahimi@ontariotechu.ca

**ABSTRACT**

Over the past few years deep learning-based techniques such as Generative Adversarial Networks (GANs) have significantly improved solutions to image super-resolution and image-to-image translation problems. In this paper, we propose a solution to the joint problem of image super-resolution and multi-modality image-to-image translation. The problem can be stated as the recovery of a high-resolution image in a modality, given a low-resolution observation of the same image in an alternative modality.
Our paper offers two models to address this problem and will be evaluated on the recovery of high-resolution day images given low-resolution night images of the same scene. Promising qualitative and quantitative results will be presented for each model.

## 1. INTRODUCTION

Super-Resolution (SR) refers to the process of creating High-Resolution (HR) images from a corresponding Low-Resolution (LR) image. In computer vision, image resolution enhancement or super-resolution has been explored for decades, yet it remains a challenging task (Wang, Z. et al, 2020).

In image-to-image translation, images are transferred from a source domain or modality to a target domain while maintaining their content representations. Its wide range of applications in many computer vision and image processing problems, such as image synthesis, segmentation, style transfer, restoration, and pose estimation have attracted increasing attention and tremendous progress in recent years (Pang, Y. et al, 2021), (Choi, Y. et al, 2020), (Fu, Y. et al, 2021).

The generative adversarial networks (GANs) are deep learning-based generative models for both semi-supervised as well as unsupervised learning that outperform state-of-the-art methods for many image-to-image translation and super-resolution tasks (Wang, Z. et al, 2020), (Alotaibi, A., 2020).

Translation and up-sampling of images simultaneously would be an interesting challenge. It is important that any model should produce realistic results in both tasks. An example would be to produce a high-resolution daytime image of a scene given a low-resolution nighttime image of the same scene. When combining super-resolution with image-to-image translation, it can be more difficult to create a realistic image, since more unknown quantities have to be recovered.

Our paper presents GAN-based solutions to the problem of combining image-to-image translation with image super-resolution. The problem is addressed via two models using Real-ESRGAN (Wang, X. et al, 2021) for image super-resolution and CUT (Park, T. et al, 2020) for image-to-image translation. The models will be evaluated both qualitatively and quantitatively.

## 2. RELATED WORK

To best of our knowledge, no studies have been conducted on the use of GANs to combine image super-resolution with image-to-image translation.

The idea of image-to-image translation dates back to Hertzmann et al.'s image analogies (Hertzmann, A. et al, 2001), a non-parametric model that uses a pair of images to generate image transformations (Zhu, J.Y. et al, 2017). There are many problems relating to computer vision and computer graphics applications that are instances of the image-to-image translation problem. Image-to-image translation involves mapping of an image given in a domain to a copy of the image in a different target domain. An example of the translation problem could be the mapping of grayscale images to RGB. In order to learn how to map one visual representation to another, it is necessary to understand the underlying features that are shared among these representations, such features can be either domain independent or domain-specific (Pang, Y. et al, 2021).

During translation, domain-independent features represent the underlying spatial structure and should be preserved (e.g., the content should be preserved when translating a natural image to Van Gogh's style), whereas domain-specific features relate to the rendering of the structure and may need to be changed during translation (e.g., if the style needs to be changed when translating the image to Van Gogh's style) (Alotaibi, A., 2020).

It is challenging to learn the mapping between two or more domains. It is sometimes difficult to collect a pair of images or the relevant images may not exist. Another challenge is where one input image can be mapped to multiple outputs. The development and use of GANs and their variants in image-to-image translation have provided state-of-the-art solutions in recent years (Alotaibi, A., 2020), (Park, T. et al, 2020), (Emami, H. et al, 2020).

Over the past few decades, researchers have been exploring image super-resolution, a technique for reconstructing images from low-resolution observations into higher resolution images. In particular, deep learning-based SR approaches have attracted much attention and have significantly improved reconstruction accuracy on synthetic data (Bashir, S.M.A. et al, 2021). GANs can be used to enhance lower quality images by reducing noise and enhancing sharpness and contrast along with the resolution (Gupta, R. et al, 2020), (Nazeri, K. et al, 2019). A super-resolution GAN combines a deep network with an adversary network to create higher resolution images (Wang, X. et al, 2021).

## 3. METHODOLOGY

The problem description and a possible methodology are presented in this Section.

### 3.1 Problem description

It is possible to formalize the task of combining super-resolution and image-to-image translation in the following way. While this paper presents the problem in the context of daytime and nighttime images, our approach can be generalized to any arbitrary multi-modality super-resolution scenario.

A LR input image in nighttime mode of a scene $N_{LR}$ is provided to the model. The model will generate a corresponding HR daytime image of the same scene $D_{HR}$ such that A. Its contents are meaningful and coherent with $N_{LR}$ so that human eye would accept it as a HR daytime image of $N_{LR}$. B. It is similar to ground truth image $D_{HR_{gt}}$, when it is available. C. Based on a daytime image, it has a realistic-looking texture and color spectrum.

### 3.2 System overview

To address this problem, we propose and investigate two different two-stage models based on GANs.

#### 3.2.1 M1: SR-first model

First stage in this model would be to generate $N_{HR}$ using Real-ESRGAN (Wang, X. et al, 2021). $N_{HR}$ is the generated HR version of $N_{LR}$. Then, CUT (Park, T. et al, 2020), which is a newer version of CycleGAN (Zhu, J.Y. et al, 2017), is fed by NHR as an input image. $N_{HR}$ would be translated to HR daytime image $D_{HR}$ in the second stage. Both Real-ESRGAN and CUT are GAN-based models and have to be trained and tuned. This approach is depicted in Figure 1.

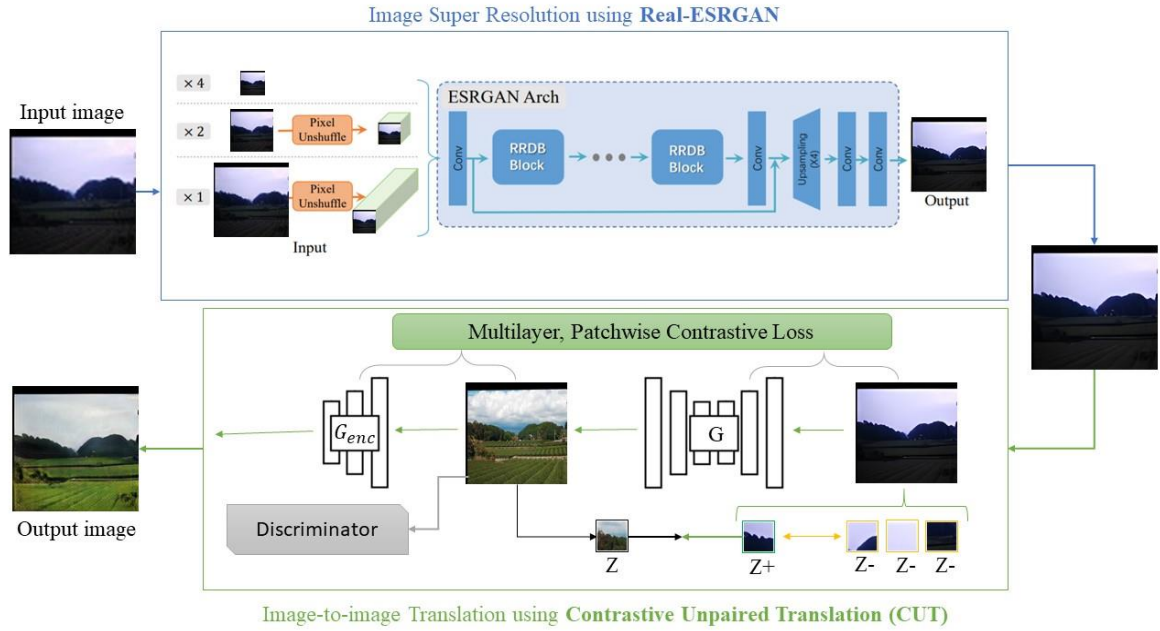

Figure 1. SR-first Model (M1); Super-resolution module followed by an image-to-image translation module.

### 3.2.2 M2: Translation-first model

This model examines the opposite order of the operations of the SR-first model. It means that the CUT (Park, T. et al, 2020) is used in the first stage to translate $N_{LR}$ into LR daytime image $D_{LR}$ followed by the second stage which is using the Real-ESRGAN (Wang, X. et al, 2021) to generate HR version of daytime image $D_{HR}$. Dimensions of input images and dimensions of output images in the CUT (Park, T. et al, 2020) are the same. Therefore, the zooming factor could be set in Real-ESRGAN (Wang, X. et al, 2021). Figure 2 illustrates how this approach works.

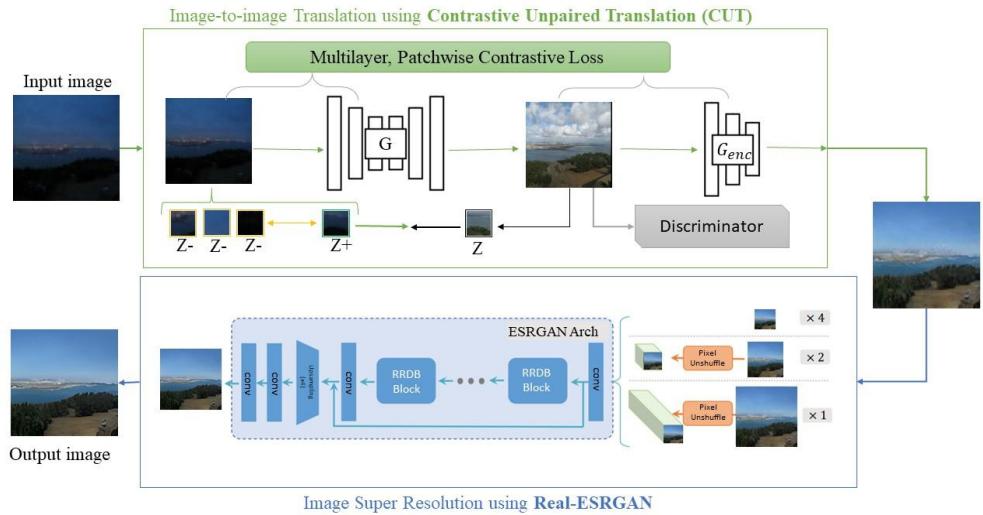

Figure 2. Translation-first Model (M2); An image-to-image translation module followed by a super-resolution module.

## 4. EXPERIMENTS AND RESULTS

In this Section, the proposed models are assessed quantitatively and qualitatively. We will also explain the dataset and setup.

### 4.1 Dataset

The Transient Attributes dataset (Laffont, P.Y. et al, 2014) was used to create the night2day dataset. Each original sample is of size 256 × 512 containing two 256 × 256 images, i.e., one day image along with its corresponding night image. These images have been divided into 20,110 night images and 20,110 corresponding day images. The dataset was then randomly split into training, validation, and test subsets for each night and day categories. More precisely, 2,011 or 10% for testing, 2,011 or 10% for validation, and 16,088 or 80% for training phases.

In SR, down-sampling is the basic operation for synthesizing LR images. In general, we consider both down-sampling and up-sampling as resizing. In addition to the nearest neighbor interpolation algorithm, there are also bilinear interpolation and bicubic interpolation algorithms. A resize operation can produce a variety of effects; some produce blurry results, whereas others may produce overly-sharp images.

In this study, the bilinear interpolation is used to convert dataset into LR images. To resize an image, OpenCV's "cv2.resize()" function can be used. This function which uses the bilinear interpolation (INTER LINEAR) as its default interpolation method, is used to produce low-resolution version (64×64) of each sample data.

Considering zooming factor of 4 in each direction, after creating a low-resolution version of each image, we have 80% or 16,088 LR night images (64×64) and 80% or 16,088 LR day images (64 × 64) for training, 10% or 2,011 LR night images (64×64) and 10% or 2,011 LR day (64×64) images for validation, and 10% or 2,011 LR night images (64×64) and 10% or 2,011 LR day images (64×64) for testing. The nearest-neighbor interpolation was used to up-sample the LR images when comparing it to the HR image.

### 4.2 Training

There are two stages of training for each model. To start the training for the first stage of the *M1 model*, a pre-trained Real-ESRGAN (Wang, X. et al, 2021) was used. Real-ESRGAN uses only HR images (256×256) for training and validation. Thus, there is no need to use the resized training data and the resized validation data at this stage. The pre-trained model is trained for 300 epochs on the *night2day* dataset. It is optimized using the Adam optimizer using an initial learning rate of 0.0002.

In the second stage of the *M1 model*, we follow the setting of the CUT (Park, T. et al, 2020). For training parameters, the number of epochs is 200. The Adam optimizer (Kingma, D.P. and Ba, J., 2014) with $\beta_1 = 0.5$ and $\beta_2 = 0.999$ is used to optimize the model. The Initial learning rate for Adam optimizer is 0.0002. The type of GAN objective is LSGAN (Mao, X. et al, 2017). Size of image buffer that stores previously generated images is 50.

In the *M2 model*, first stage is translation of the LR night image of size (64×64) into the LR day image (64×64). The CUT (Park, T. et al, 2020) input image and output image sizes are changed from 256×256 to 64×64 by adjusting the scaling factor. Therefore, LR night images and LR day images are used to train the translation module. In order to keep consistency, the *CUT*'s settings from the previous approach were used. In the second stage of the *M2 model*, a trained Real-ESRGAN (Wang, X. et al, 2021) model was used with the same settings as the first stage of the *M2 model*.

We conducted all experiments using the Linux Ubuntu operating system and the NVIDIA GeForce GTX TITAN X GPU.

### 4.3 SR-first model (M1) Evaluation

A model evaluation is performed using 2,011 LR nighttime images (64×64). Three evaluation phases have been defined for a better quantitative evaluation of the M1 model.

- M1-Pre: Comparing *the LR night image (64 × 64) (Input)* VS *Real day image i.e. Day Ground truth image*.
- M1-Intermediate: Comparing the generated HR night image with 256×256 resolution (After SR) VS Real day image i.e. Day ground truth image.
- M1-Post: Comparing the generated HR day image with 256×256 resolution (Output) VS Real day image i.e. Day ground truth image.

By comparing M1-Pre with M1-Intermediate and M1Post, the overall performance of the M1 model can be assessed. Root Mean Square Error (RMSE), Mean Absolute Error (MAE), Structural Similarity Index Measure (SSIM), Normalized Cross-Correlation (NCC), and Fréchet inception distance (FID) are calculated for each phase and histograms of these similarity measures over the test data are provided. The comparisons are shown in Figure 3.

Figure 3 shows that the RMSE (top-left) and the MAE (top-right) are shifted to the left from the "M1-Pre" state to the "M1-Post" state. Lower value of these measures for final results in "M1-Post" indicates better performance. Similarly, the "M1-Post" phase provides higher values for SSIM (bottom-left), and the NCC (bottom-right) in comparison with the "M1-Intermediate" and the "M1-Pre" states. Table I presents the mean and the standard deviation of these measures for Pre, Intermediate, and Post. The FID metric for each phase is computed as well. Based on all these measures, Table 1 suggests that final results from the model are more consistent with ground truth.

Table 1. Comparing final results M1-Pre vs M1-Intermediate vs M1-Post of the SR-first model (M1) using different measures.

|   | M1-Pre | M1-Intermediate | M1-Post |
| --- | --- | --- | --- |
| RMSE (↓) | 0.48±0.14 | 0.36±0.10 | **0.23±0.08** |
| MAE (↓) | 0.43±0.14 | 0.30±0.09 | **0.06±0.04** |
| SSIM (↑) | 0.34±0.14 | 0.36±0.14 | **0.43±0.14** |
| NCC (↑) | 0.81±0.14 | 0.83±0.13 | **0.91±0.08** |
| FID (↓) | 173.95 | 146.76 | **90.47** |

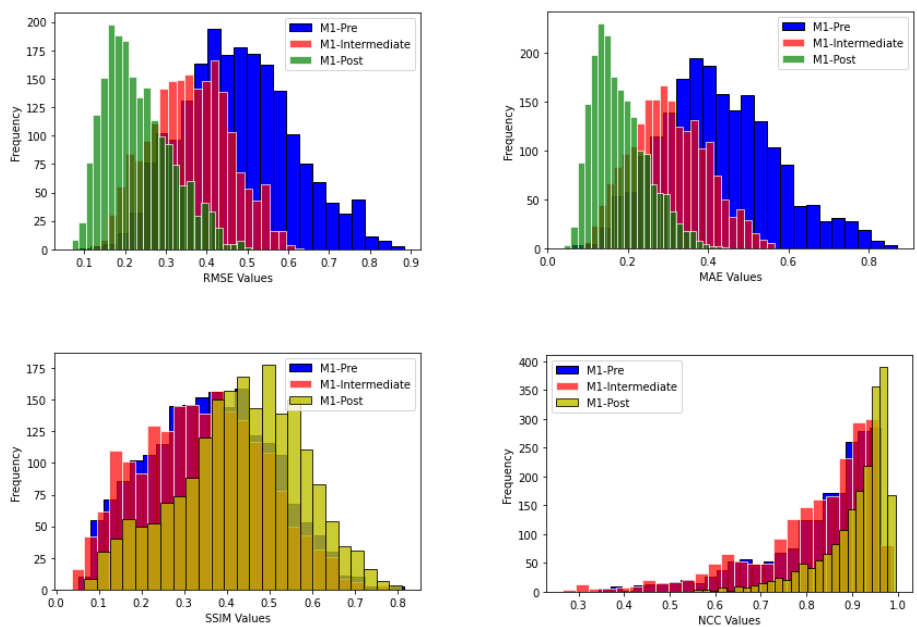

Figure 3. Comparison of final results (Pre vs Intermediate vs Post) of the SR-first model (M1) using different measures.

Some examples of images for different stages of the M1 model is provided in Figure 4 for qualitative evaluation. From left to right, each column shows the following: a) Input image (LR night image), b) The generated HR night image (after super-resolution), c) Real night image (night ground truth image), d) Output (generated HR day), and e) Real day image (day ground truth image).

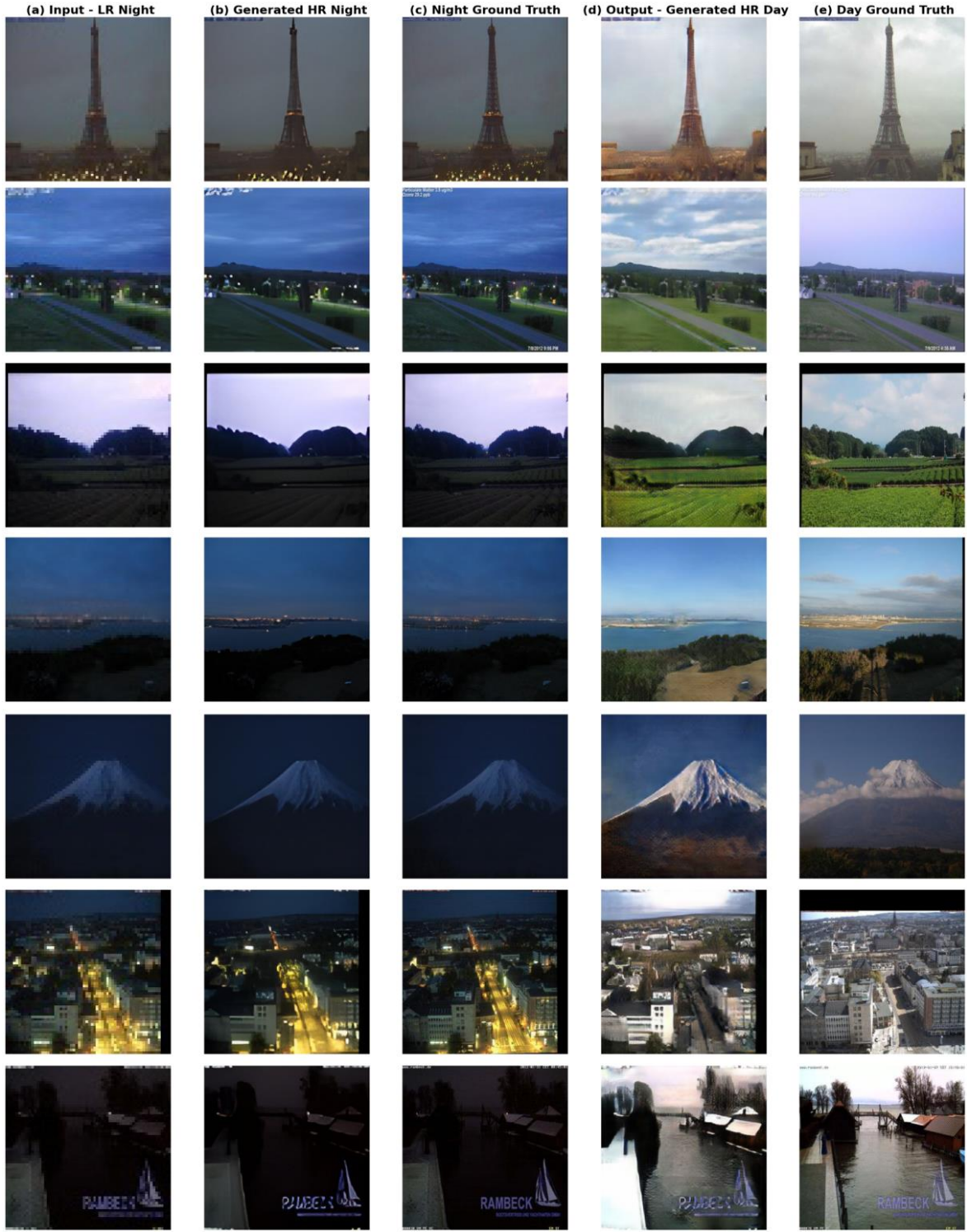

Figure 4. Qualitative evaluation of M1 model; from left to right; (a) Input image, (b) Generated HR night image, (c) Night ground truth, (d) Output image, (e) Day ground truth.

## 4.4 Translation-first model (M2) Evaluation

LR nighttime images (64×64) are used to evaluate the model. For a more quantitative analysis of the M2 model, two phases have been defined.

• M2-Pre: Comparing the LR night image (64 × 64) (Input) VS Real day image i.e. Day ground truth image.

• M2-Post: Comparing the generated HR day image with 256×256 resolution (Output) VS Real day image i.e. Day ground truth image.

In addition, M2-Intermediate phase could have also been defined as comparing the generated LR day image (64×64) (after translation) VS Real day image i.e. day ground truth image. However, the output size of the translation module in the M2 model would be different from that of the first stage if it had been up-sampled. As a result, this evaluation phase was not considered and the total performance of the M2 model is evaluated only by comparing M2-Pre with M2-Post.

For RMSE, MAE, SSIM, NCC, and FID, we compare the results of these evaluation phases. Figure 5 illustrates these comparisons.

In Figure 5, it can be seen that the RMSE (top-left) and the MAE (top-right) are shifted to the left significantly from the "Pre" to the "Post". Therefore, lower values of these measures for final results in M2-Post indicate better performance. The SSIM (bottom-left) and NCC (bottom-right) are shifted to the right in the "Post", which means that the generated images are more similar to the desired output.

In Table 2, these measures are listed along with their means and standard deviations for M2-Pre, and M2-Post. Furthermore, FID metrics for each phase are calculated. There is a ↓ sign next to each measure indicates that a lower value is "better", and similarly, a ↑ sign means a higher value is "better".

Some examples of images for different stages of the M2 model is provided in Figure 6 for qualitative evaluation. From left to right, each column shows the following: a) Input image (LR night image), b) The generated LR day image (after translation), c) Output (generated HR day), and d) Real day image (day ground truth image).

Table 2. Comparing final results M2-Pre vs M2-Post of the Translation-first model (M2) using different measures.

|  | M2-Pre | M2-Post |
| --- | --- | --- |
| RMSE (↓) | 0.48±0.14 | **0.24±0.08** |
| MAE (↓) | 0.43±0.14 | **0.07±0.07** |
| SSIM (↑) | 0.34±0.14 | **0.44±0.14** |
| NCC (↑) | 0.81±0.14 | **0.90±0.08** |
| FID (↓) | 173.95 | **96.56** |

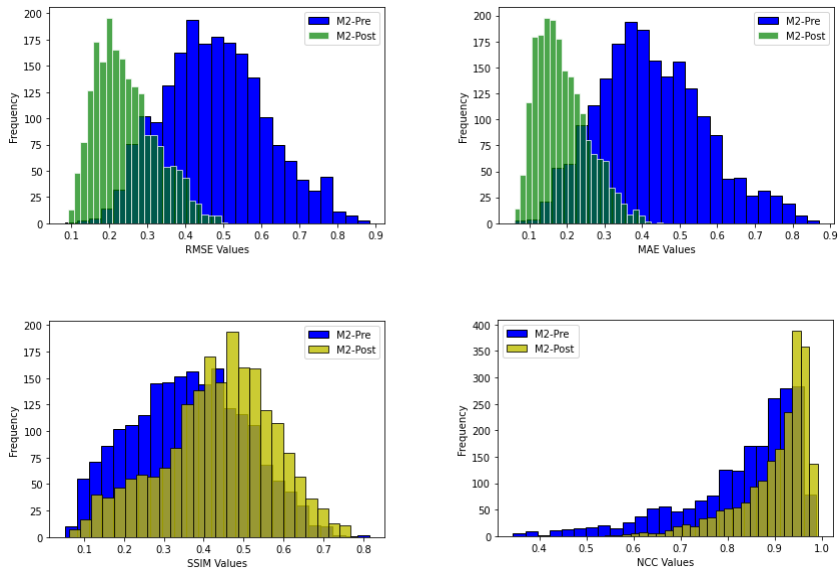

Figure 5. Comparison of final results (Pre vs Post) of the Translation-first model (M2) using different measures.

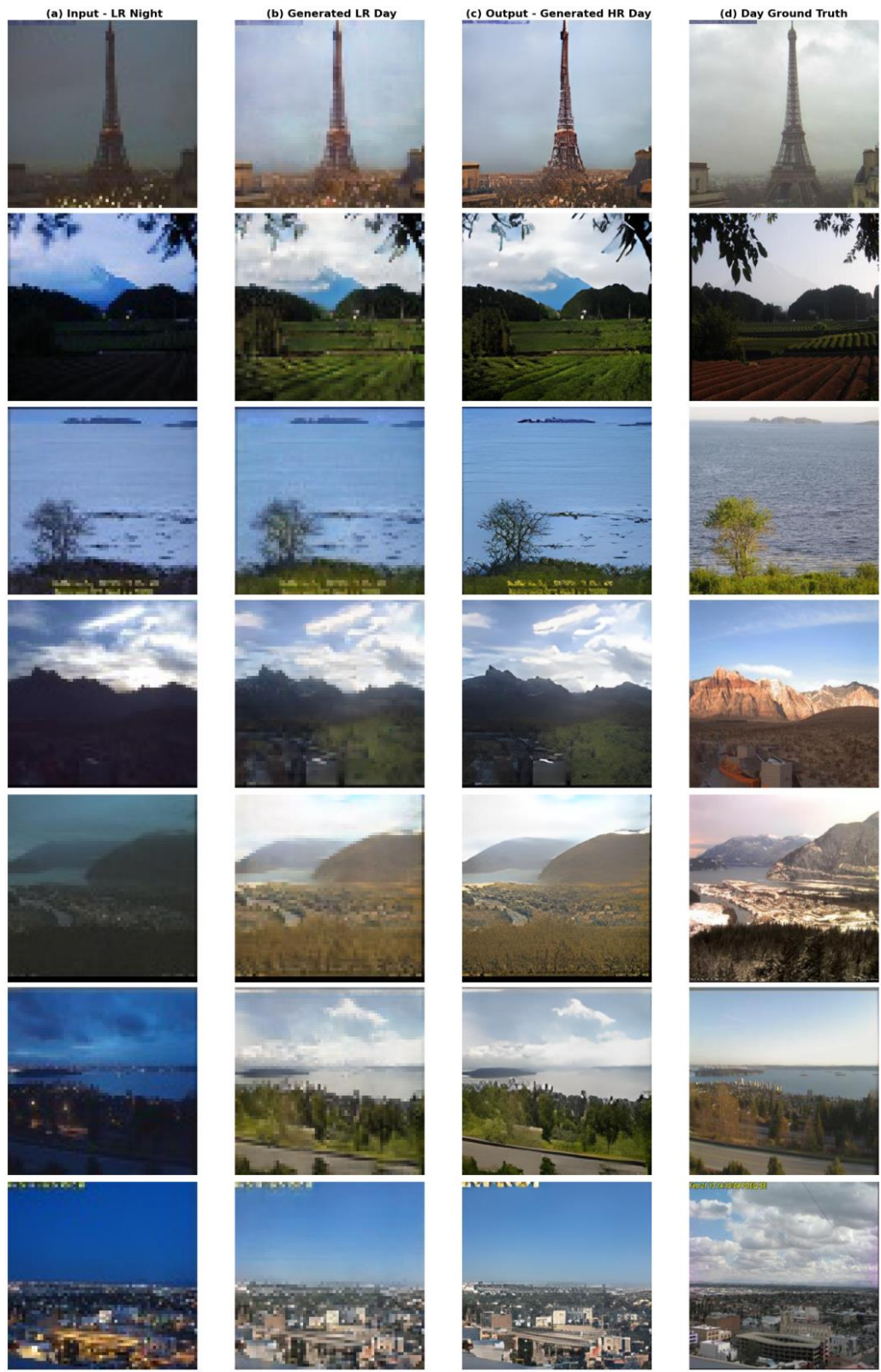

Figure 6. Qualitative evaluation of the translation-first model (M2) model; from left to right; (a) Input (LR night image), (b) The generated LR day (after translation), (c) Output (generated HR day), and (d) Real day image (day ground truth)

## 4.5 SR-first model (M1) VS Translation-first model (M2)

Comparing the **M1-Post** and **M2-Post** results will enable us to compare the M1 and M2 models' performance. In both models, the Post phase is comparing the generated HR day image (Output) VS Real day image or day ground truth image. This comparison using the RMSE (top-left), the MAE (top-right), the SSIM (bottom-left), and the NCC (bottom-right) is depicted in Figure 7. Table 3 summarizes these comparisons.

Table 3. Comparing final results of the **M1: SR-first** with **M2: Translation-first** models using different measures.

|  | M2 | M1 |
|---|---|---|
| RMSE (↓) | 0.24±0.08 | **0.23±0.08** |
| MAE (↓) | 0.07±0.07 | **0.06±0.04** |
| SSIM (↑) | **0.44±0.14** | 0.43±0.14 |
| NCC (↑) | 0.90±0.08 | **0.91±0.08** |
| FID (↓) | 96.56 | **90.47** |

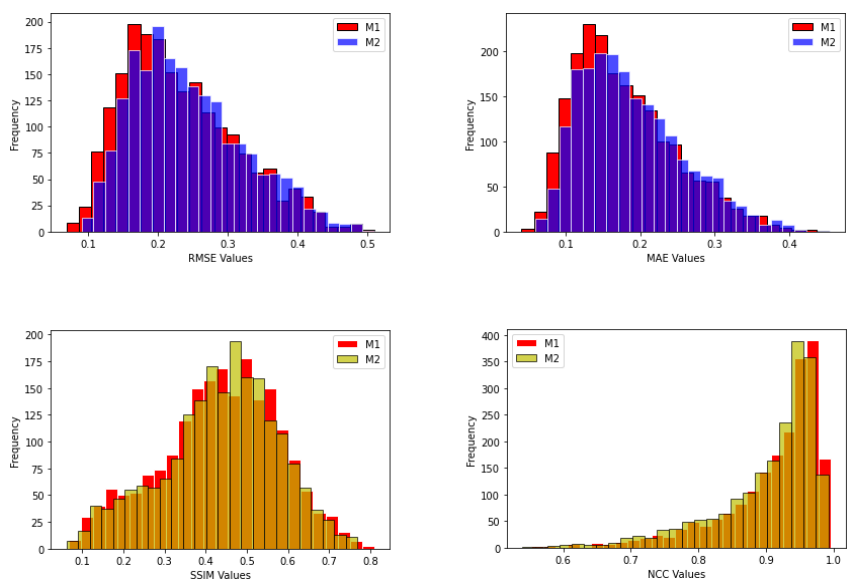

Figure 7. Comparison of final results of the **M1: SR-first** with **M2: Translation-first** models using different metrics.

The models can be assessed qualitatively using figures 4 and 6. The quantitative and qualitative results indicate that the two models perform similarly and that there is no significant difference between them.

## 5. CONCLUSION

The problem of combining the image-to-image translation with the image super-resolution was proposed and addressed by concatenating two existing modules and rigorously evaluating the performance of both models after training. This paper describes the steps involved in preparing the selected dataset for the defined problem. In addition, a variety of evaluation phases are used to evaluate M1 and M2 models. Qualitative and quantitative analysis indicate that both models perform similarly.

By performing both tasks simultaneously, an end-to-end generative network may provide more accurate results. Our study focuses on the night-to-day image super-resolution and provides promising results.

In many image enhancement tasks including medical imaging, a similar approach can be used to generates realistic high-resolution image of a desired modality based on a given low resolution image in a different modality.

## ACKNOWLDGEMENTS

This work was supported in part by the Natural Sciences and Engineering Research Council of Canada (NSERC).